\documentclass[12pt]{article}

\usepackage{amsmath}
\usepackage[dvips]{graphicx}

\def\os{\overline{s}}
\def\oa{\overline{a}}
\def\oc{\overline{c}}
\def\od{\overline{d}}

\begin{document}

\begin{flushright}
March, 2008
\end{flushright}

\vspace{0.0cm}

\begin{center}

{\Large \bf Twisted SUSY Invariant Formulation of \\
\vspace{0.3cm}

Chern-Simons Gauge Theory on a Lattice}

\vspace{1cm}

{ Kazuhiro Nagata}\footnote{knagata@indiana.edu}$^{*}$ and
{ Yong-Shi Wu}\footnote{wu@physics.utah.edu}$^{\dagger}$\\

\vspace{0.5cm}
$^{*}${\it{ Department of Physics, Indiana University}}\\
{\it{ Bloomington, IN 47405, USA}}\\
$^{\dagger}${\it{ Department of Physics, University of Utah }}\\
{\it{ Salt Lake City, UT 84112, USA}}
\end{center}

\begin{abstract}
We propose a twisted SUSY invariant formulation of Chern-Simons
theory on a Euclidean three dimensional lattice. The SUSY algebra to be
realized on the lattice is the $N=4\ D=3$ twisted algebra that was
recently proposed by D'Adda {\it et al}. 
In order to keep the manifest
anti-hermiticity of the action, we introduce oppositely oriented
supercharges. Accordingly, the naive continuum limit of the action
formally corresponds to the Landau gauge fixed version of Chern-Simons
theory with complex gauge group which was originally proposed
by Witten. 
We also show that the resulting action consists of parity even
and odd parts with different coefficients.

\end{abstract}

\section{Introduction}
\indent

Chern-Simons gauge theory is a fundamentally important field
theory in both physics and mathematics. The Lagrangian density
\cite{Siegel79,Jackiw82} is just the famous Chern-Simons secondary
characteristic class \cite{CS74} for a principal bundle. As a
topological field theory, its action can be defined in an odd
dimensional spacetime without involving its metric. So
mathematically the metric independent physical observables of the
theory are topological invariants independent of spacetime metric
\cite{Witten89}. In particular, the partition function of the
theory on a compact manifold gives rise to a three-manifold
invariant, while the expectation value of Wilson loops gives rise
to knot-link invariants, say Jones polynomials \cite{Jones85} in
the case with the non-Abelian gauge group $SU(2)$. On the physics
side, by now it is well-known that Chern-Simons gauge theory can
be used as low energy effective theory to describe a new type of
matter, the so-called topological phases, in planar condensed
matter systems (or in 2+1 spacetime dimensions), such as the
Fractional Quantum Hall effect \cite{Wen94}. Also quantum gravity
in (2+1)-dimensional spacetime, which is known to be
diffeomorphism invariant, can be formulated as a Chern-Simons
theory with the Poincar\'e group as gauge group \cite{Witten88}.
In recent years, the close relationship between Chern-Simons gauge
theory, topological invariants and topological phases has
attracted a lot of attention for developing topological quantum
computation \cite{Freedman03,DasSarma07}. The above is just only a
few examples of the ubiquitousness of the Chern-Simons theory in
physical applications. For a recent survey see, e.g.,
ref.\cite{Jackiw05}.

Because of the primary importance of Chern-Simons theory, it is
much desirable to put the theory on a lattice for the convenience
of computer simulations. However, up to now this task has been
achieved with a limited success. Previously lattice formulations
of Chern-Simon theory have been addressed in the context of
bosonization or anyonization \cite{Frohlich,Luscher,Kantor} or of
topological excitations \cite{Diamantini} in a regularized
framework.\footnote{There are works on simplicial lattices
addressing abelian Chern-Simons theory
in terms of geometric discretization scheme \cite{Sen:2000ez}
and also Chern-Simons gravity via Ponzano-Regge model \cite{Kawamoto:1999tf}.}
Two major difficulties in formulating lattice
Chern-Simons theory have been identified. One is the problem of an
extra zero-eigenvalue in the gauge field kernel, which arises from
the fact that the gauge kinetic terms involve only first order
derivatives. This feature resembles the ``doubling problem" for
lattice fermions, which is also tightly connected with the
hermiticity issue of the lattice action \cite{Berruto,Fosco}. The
other difficulty, in formulating a non-Abelian Chern-Simons on a
lattice, is related to gauge non-invariance of the action for a
non-Abelian theory under large gauge transformations.

In this paper we attack the problem of the lattice formulation of
Chern-Simons theory with a new method. Instead of attempting to
directly put the Chern-Simons action on a lattice, we propose to
put the gauge fixed Chern-Simons theory on a Euclidean lattice. We
also introduce oppositely oriented component fields in order to
ensure the manifest anti-hermiticity of the lattice action. We are
motivated by two observations in the literature. The first
observation is an old one \cite{Birmingham:1988ap,Birmingham:1989cs},
that there exists a very rich symmetry structure in the Landau-gauge fixed
Chern-Simons action; namely, apart from the ordinary BRST
symmetry, there exist more fermionic symmetries of vector type. In
ref. \cite{DGS}, the set of symmetries together with the anti-BRST
type symmetries are identified as a certain type of twisted
supersymmetry (SUSY), which was originally proposed in the context
of topological quantum field theory \cite{Witten:1988ze}. Since
then, the twisted SUSY invariant properties of the Chern-Simons
theory in Landau gauge have been studied in more detail concerning
its quantum aspects \cite{Delduc:1990je} 
as well as its rich symmetry structure
\cite{Damgaard:1990wm}. 
The second observation that inspires us is a recent one, that the
twisted SUSY plays a particularly important role in realizing SUSY
on a lattice \cite{DKKN1,DKKN2,DKKN3,Nagata,twist-lattsusy}. This is
essentially due to the intrinsic relation between twisted fermions
and Dirac-K\"ahler fermions \cite{KT}. It is observed that among
other recent developments of lattice SUSY \cite{other-lattsusy1},
the so-called deconstruction formulation of lattice SUSY
\cite{deconstruction} can also be related to the twisted SUSY
framework \cite{relation}. Motivated by these recent developments,
we naturally anticipate that a lattice formulation of Chern-Simons
theory can be given through the lattice realization of the twisted
SUSY associated with the Landau gauge fixed action.

This article is devoted to constructing a Landau gauge fixed
Chern-Simons multiplet directly on a three dimensional lattice and
to proposing a manifestly anti-hermitian Euclidean lattice action.
This paper is organized as follows. In Sec. 2, we review the
symmetries of the Landau gauge fixed Chern-Simons action in
continuum spacetime. In Sec. 3, after giving an overview of the
twisted SUSY formulation on a lattice developed in \cite{DKKN1}
and introducing the twisted $N=4\ D=3$ lattice algebra
\cite{DKKN3}, we proceed to construct a lattice counterpart of the
Chern-Simons multiplet. We also introduce oppositely oriented
supercharges and component fields in order to realize the manifest
(anti-)hermiticity of the lattice multiplet. In Sec. 4, we
construct a lattice version of Landau gauge fixed Chern-Simons
action and show how the twisted SUSY invariance is realized. We
further show that the zero-eigenvalue problem does not occur in
our formulation owing to manifest anti-hermiticity of the lattice
action. We also discuss about the naive continuum limit and its
relation to the Chern-Simons theory with complex gauge group
\cite{Witten:1989ip}. Sec. 5 addresses the parity transformation
properties of our lattice action, and Sec. 6 summarizes our
formulation with some discussions.


\section{Chern-Simons in Landau gauge}
\indent

In this section, we review the symmetry aspects of the
Chern-Simons action with Landau gauge fixing in the continuum
spacetime. Although the original Chern-Simons action is given in a
metric independent form, it becomes metric dependent after the
gauge fixing terms are introduced. In this paper, we only consider
the Euclidean three dimensional flat spacetime. The gauge fixed
action is given by
\begin{eqnarray}
S &=& i \frac{k} {2\pi} \int d^{3}x \mathrm{Tr} \biggl[
\epsilon_{\mu\nu\rho}(\frac{1}{2}A_{\mu}\partial_{\nu}A_{\rho}
+\frac{1}{3}A_{\mu}A_{\nu}A_{\rho})
-b\partial_{\mu}A_{\mu}-\overline{c}\partial_{\mu}D_{\mu}c
\biggr], \label{CSgf}
\end{eqnarray}
where $A_{\mu}$, $b$, $c$ and $\overline{c}$ denote the gauge
field, an auxiliary field, the ghost and anti-ghost field,
respectively. The coefficient $k$ should be a multiple of integer
required by invariance under large gauge transformations. Note the
the overall purely imaginary factor $i$ in the Euclidean action,
because the path integral measure of the topological field theory
has to be a pure phase factor. All of the component fields belong
to the adjoint representation of the gauge group with the
following anti-hermiticity conditions \cite{DGS},
\begin{eqnarray}
A_{\mu}^{\dagger} &=& -A_{\mu}, \hspace{20pt}
b^{\dagger} \ =\ -b, \hspace{20pt}
c^{\dagger} \ =\ -c,\hspace{20pt}
\overline{c}^{\dagger} \ =\ \overline{c}. \label{hermiticity_cont}
\end{eqnarray}
The gauge fixed action (\ref{CSgf}) is invariant under the BRST
transformations which are remnants of the original gauge symmetry,
\begin{eqnarray}
sA_{\mu} &=& -D_{\mu}c, \hspace{30pt}
sc \ =\ c^{2},  \\
s\overline{c} &=& b, \hspace{55pt}
s b \ =\ 0 ,
\end{eqnarray}
where the covariant derivative $D_{\mu}$ is defined by $D_{\mu}c =
\partial_{\mu}c+[A_{\mu},c]$. Furthermore, it was pointed out in
\cite{Birmingham:1988ap,Birmingham:1989cs,DGS} that the action
(\ref{CSgf}) is also invariant under additional fermionic
transformations including vector-type transformations,
$\overline{s}_{\mu}$, $s_{\mu}$ and $\overline{s}$, where the
index $\mu$ runs from 1 to 3. We list their transformation laws
for the component fields in Table \ref{transCScont}. The whole set
of eight generators $(s,\overline{s}_{\mu},s_{\mu},\overline{s})$
is shown to satisfy the following algebra \cite{DGS},
\begin{eqnarray}
\{s,\overline{s}_{\mu}\} &\dot{=}& \partial_{\mu}, \hspace{39pt}
\{s_{\mu},\overline{s}_{\nu}\} \ \dot{=}\
\epsilon_{\mu\nu\rho}\partial_{\rho}, \label{Alg_cont1} \\ [2pt]
\{\overline{s},s_{\mu}\} &\dot{=}& -\partial_{\mu}, \hspace{20pt}
\{others\} \ =\ 0. \label{Alg_cont2}
\end{eqnarray}
Here the dotted equality means that the algebra closes only
on-shell, namely up to equations of motion. The anti-hermiticity
conditions for the twisted supercharges can be imposed
consistently with those for the component fields
(\ref{hermiticity_cont}):
\begin{eqnarray}
s^{\dagger} &=& -s, \hspace{20pt}
\overline{s}^{\dagger} \ =\  \overline{s},\hspace{20pt}
s_{\mu}^{\dagger} \ =\ -s_{\mu}, \hspace{20pt}
\overline{s}_{\mu}^{\dagger} \ =\  \overline{s}_{\mu},
\hspace{20pt}
\partial_{\mu}^{\dagger} \ =\ -\partial_{\mu}.
\end{eqnarray}
Since the BRST generator $s$ is supposed to transform as a scalar
under the Lorentz transformation, we immediately read off from the
algebra (\ref{Alg_cont1})-(\ref{Alg_cont2}) that the remaining
fermionic generators $\overline{s}_{\mu}$, $s_{\mu}$ and
$\overline{s}$ transform as a vector, another vector and a scalar,
respectively. These transformation properties are identical to the
ones in a certain type of twisted SUSY, where the new Lorentz
group, which is called the twisted Lorentz group, is defined as
the diagonal subgroup of the original Lorentz group and a certain
type of internal symmetry group. In the present case with eight
supercharges $(s,\overline{s}_{\mu},s_{\mu},\overline{s})$, the
twisted Lorentz group is understood as the diagonal subgroup of
$SO(3)_{Lorentz}\times SO(3)_{Internal}$ whose covering group is
$(SU(2)\times SU(2))_{diag}$. The twisted structure can be
explicitly seen from the following combinations of
$(s,\overline{s}_{\mu},s_{\mu},\overline{s})$ into the generators
$Q_{\alpha k }$ and $\overline{Q}_{k \alpha}$ with spin index
$\alpha$ and internal index $k$:
\begin{eqnarray}
Q_{\alpha k} &=& (\mathbf{1}s + \gamma_{\mu}(-is_{\mu}))_{\alpha k}, \\
\overline{Q}_{k \alpha} &=& (\mathbf{1}\overline{s} +
\gamma_{\mu}(i\overline{s}_{\mu}))_{k \alpha},
\end{eqnarray}
where $\mathbf{1}$ represents the unit matrix while $\gamma_{\mu}
(\mu=1,2,3)$ represent three dimensional gamma matrices which can
be taken to be Pauli matrices. One can easily see that $s_{\mu}$
and $\overline{s}_{\mu}$ transform as vector if the spin and
internal indices are rotated simultaneously. Furthermore, in terms
of $Q_{\alpha k }$ and $\overline{Q}_{k \alpha}$, the algebra
(\ref{Alg_cont1})-(\ref{Alg_cont2}) can be re-expressed as
\begin{eqnarray}
\{Q_{\alpha k},\overline{Q}_{l \beta}\} &=&
2i\delta_{kl}(\gamma_{\mu})_{\alpha\beta}
\partial_{\mu}.
\end{eqnarray}
This clearly shows that the internal symmetry indices $k$ and $l$
can be viewed as the suffices labeling extended SUSY, while
$\alpha$ and $\beta$ remain the ordinary spinor indices. From the
above observations it becomes clear that the fermionic symmetries
associated with the Landau gauge fixed Chern-Simons action are
essentially connected with a certain type of extended SUSY through
the twisting procedure. Following the standard nomenclature in
topological field theory \cite{Blau:1996bx,Geyer:2001yc}, we refer
to the algebra (\ref{Alg_cont1})-(\ref{Alg_cont2}) as the $N=4\
D=3$ twisted SUSY algebra.\footnote{In the early literatures it
was referred to as $N=2$ algebra.} A superfield formulation
based on the twisted $N=4\ D=3$ SUSY algebra is recently
elaborated in Ref.\cite{Nagata} with a direct application to
continuum super Yang-Mills theories in the off-shell regime.

\begin{table}
\begin{center}
\renewcommand{\arraystretch}{1.4}
\renewcommand{\tabcolsep}{10pt}
\begin{tabular}{c|cccc}
\hline
& $s$ & $\overline{s}_{\rho}$ & $s_{\rho}$ & $\overline{s}$ \\ \hline
$c$ & $c^{2}$ & $-A_{\rho}$ & $0$ & $-b+\{\overline{c},c\}$\\
$\overline{c}$ & $b$ & $0$ & $A_{\rho}$ & $\overline{c}^{2}$ \\
$A_{\mu}$ & $-D_{\mu}c$ & $-\epsilon_{\rho\mu\nu}\partial_{\nu}\overline{c}$ &
$-\epsilon_{\rho\mu\nu}\partial_{\nu}c$ & $-D_{\mu}\overline{c}$ \\
$b$ & $0$ & $\partial_{\rho}\overline{c}$ & $D_{\rho}c$ & $[\overline{c},b]$ \\ \hline
\end{tabular}
\caption{Fermionic transformation laws in continuum spacetime}
\label{transCScont}
\end{center}
\end{table}

It is important to mention here about parity transformations
of the component fields and the supercharges of the Chern-Simons multiplet.
Since we are working on a Euclidean three dimensional spacetime,
a parity operation on the spacetime coordinates may be defined by
the simultaneous inversion of all the directions,
\begin{eqnarray}
P (x_{1},x_{2},x_{3}) P^{-1} = (-x_{1},-x_{2},-x_{3}). \label{def_parity}
\end{eqnarray}
The gauge fields and the derivative operators are supposed to
transform as vectors, obeying
\begin{eqnarray}
P A_{\mu}(x) P^{-1} &=& -A_{\mu}(-x), \hspace{30pt}
P \partial_{\mu} P^{-1} \ =\ -\partial_{\mu},
\end{eqnarray}
where $-x$ denotes $-x=(-x_{1},-x_{2},-x_{3})$. The parity nature
of the supercharges could be determined consistently with the SUSY
transformations of the component fields, provided parity is
compatible with the SUSY algebra
(\ref{Alg_cont1})-(\ref{Alg_cont2}). Here we assume that the ghost
field $c(x)$ transforms as a scalar, namely $P c(x) P^{-1}=
c(-x)$. We then immediately read off the parity of the
supercharges as
\begin{eqnarray}
P s P^{-1} &=& s, \hspace{20pt}
P \overline{s}_{\mu} P^{-1} \ = \ -\overline{s}_{\mu}, \hspace{20pt}
P s_{\mu} P^{-1} \ =\ s_{\mu}, \hspace{20pt}
P \overline{s} P^{-1} \ = \ -\overline{s}.
\end{eqnarray}
The parity of $\overline{c}$ and $b$ are accordingly given by
$P\overline{c}(x)P^{-1}= -\overline{c}(-x)$ and $P b(x) P^{-1} =
-b(-x)$. Notice that the entire action (\ref{CSgf}) is parity odd
under these assumptions.


\section{Twisted SUSY \& Chern-Simons Multiplet on Lattice}

\subsection{Lattice SUSY algebra}

\indent

It was recently recognized \cite{DKKN3} that the $N=4$ $D=3$
twisted SUSY algebra could be realized on a lattice consistently
with the lattice Leibniz rule; then it was immediately applied to
a twisted super Yang-Mills formulation on a three dimensional
lattice. We first briefly review the lattice formulation of the
twisted SUSY proposed in ref. \cite{DKKN1} and then proceed to
construct the Chern-Simons multiplet based on the $N=4\ D=3$
twisted SUSY structure on the lattice. Since the lattice spacing
is always finite, on a lattice all the derivative operators should
be replaced by difference operators:
\begin{eqnarray}
\partial_{\mu} &\rightarrow& \Delta_{\pm\mu},
\end{eqnarray}
where $\pm$ denotes forward and backward difference, respectively.
The operation of difference on a function $\Phi(x)$ is defined
by the following type of ``shifted" commutators,
\begin{eqnarray}
(\Delta_{\pm\mu}\Phi(x)) &\equiv& \Delta_{\pm}\Phi(x) -
\Phi(x\pm n_{\mu})\Delta_{\pm\mu},
\label{diff}
\end{eqnarray}
where $n_{\mu}\ (\mu=1,\cdots,r)$ denote the unit vectors in $r$
dimensions, whose component is given by
$(n_{\mu})_{\rho}=\delta_{\mu\rho}$. We take the lattice spacing
to be unity. The difference operators $\Delta_{\pm\mu}$ are most
naturally located on links from $x$ to $x\pm n_{\mu}$ for generic
value of $x$, and they take unit values such that the definition
(\ref{diff}) actually gives the forward and backward difference:
\begin{eqnarray}
\Delta_{\pm\mu} &=& (\Delta_{\pm\mu})_{x\pm n_{\mu},x} \
=\ \mp 1. \label{delta=1}
\end{eqnarray}
Starting from the definition $(\ref{diff})$, one finds that the
operation of $\Delta_{\pm\mu}$ on a product of functions
$\Phi_{1}(x)\Phi_{2}(x)$ gives
\begin{eqnarray}
(\Delta_{\pm\mu}\Phi_{1}(x)\Phi_{2}(x))
&=& (\Delta_{\pm\mu}\Phi_{1}(x))\Phi_{2}(x)
+\Phi_{1}(x\pm n_{\mu})(\Delta_{\pm\mu}\Phi_{2}(x)),
\end{eqnarray}
which we refer to as the Leibniz rule on the lattice. The
importance of the Leibniz rule has also been recognized in the
context of non-commutative differential geometry on a lattice
\cite{KK}. Since in continuum, SUSY is essentially nothing but the
fermionic decomposition of the differential operators
$\partial_{\mu}$, we may then naturally expect that the fermionic
decomposition of the difference operators $\Delta_{\pm\mu}$ will
accordingly serve as the starting point of a lattice formulation
of SUSY. In order to be compatible with the link nature of
difference operators, we introduce a generic lattice supercharge
$Q_{A}$ on a link from $x$ to $x+a_{A}$:
\begin{eqnarray}
Q_{A} &=& (Q_{A})_{x+a_{A},x}\,\,\, ,
\end{eqnarray}
where the $a_{A}$ denotes a generic vector
whose expression is to be determined in the following.
The operation of $Q_{A}$
is again defined as a ``shifted" (anti-)commutator,\footnote{We thank
A. Jourjine for his comment on the ``shifted" (anti-)commutator
from the cell-complex cohomological point of view
and for letting us know his works \cite{Jourjine:1986hh}.
For recent works on algebraic topology 
in connection with Dirac-K\"ahler fermion on a lattice,
one may also refer to Ref. \cite{deBeauce:2005ny}.}
\begin{eqnarray}
(Q_{A}\Phi(x)) &\equiv& (Q_{A})_{x+a_{A},x}\Phi(x)
-(-1)^{|\Phi|}\Phi(x+a_{A})(Q_{A})_{x+a_{A},x}. \label{Qf}
\end{eqnarray}
Accordingly, the operation on a product of functions gives
\begin{eqnarray}
(Q_{A}\Phi_{1}(x)\Phi_{2}(x))
&=& (Q_{A}\Phi_{1}(x))\Phi_{2}(x)
+(-1)^{|\Phi_{1}|}\Phi_{1}(x+a_{A})(Q_{A}\Phi_{2}(x)), \label{Qfg}
\end{eqnarray}
where $|\Phi|$ stands for $0$ or $1$ for bosonic or fermionic
$\Phi$, respectively. The anti-commutator of these supercharges
may naturally be defined as the successive connections of link
operators:
\begin{eqnarray}
\{Q_{A},Q_{B}\}_{x+a_{A}+a_{B},x} &\equiv&
(Q_{A})_{x+a_{A}+a_{B},x+a_{B}}(Q_{B})_{x+a_{B},x} +
(Q_{B})_{x+a_{A}+a_{B},x+a_{A}}(Q_{A})_{x+a_{A},x}. \qquad
\end{eqnarray}

\begin{figure}
\begin{center}
\begin{minipage}{60mm}
\begin{center}
\includegraphics[width=50mm]{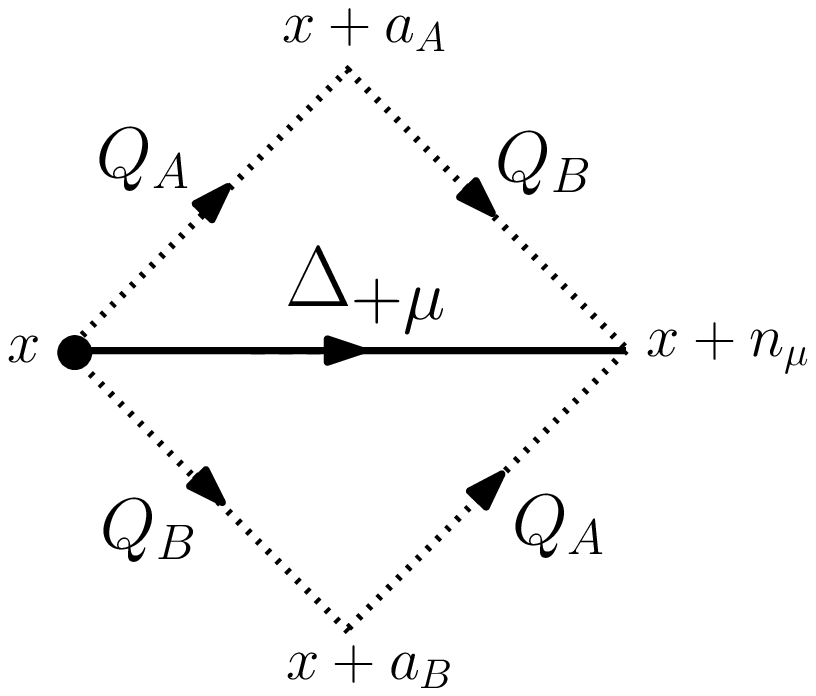}
\caption{Lattice SUSY algebra subject to the condition (\ref{alg_forward})}
\label{s_Delta1}
\end{center}
\end{minipage}
\hspace{20pt}
\begin{minipage}{60mm}
\begin{center}
\includegraphics[width=50mm]{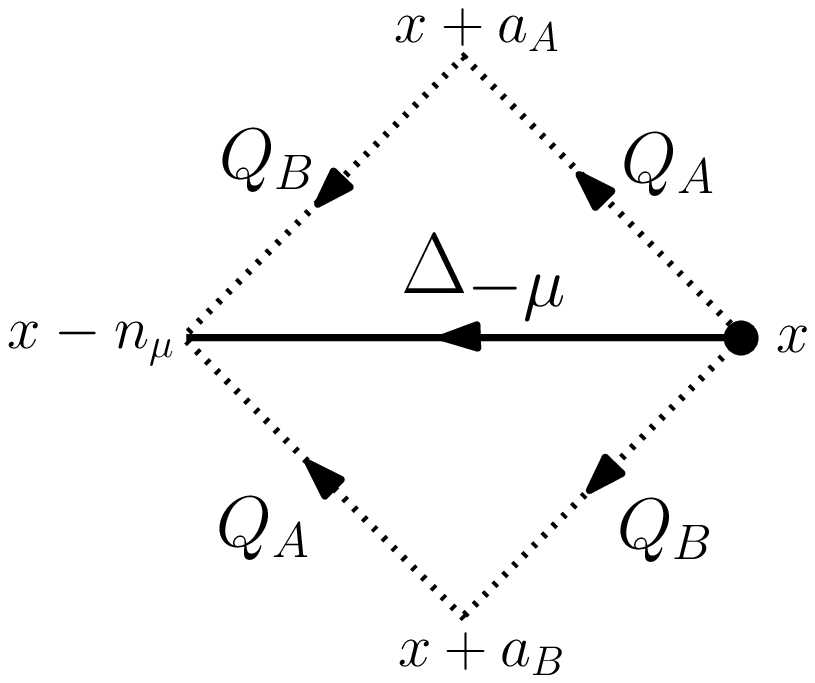}
\caption{Lattice SUSY algebra subject to the condition (\ref{alg_backward})}
\label{s_Delta2}
\end{center}
\end{minipage}
\end{center}
\end{figure}

In terms of the above link operators, we can express the generic
form of lattice SUSY algebra as
\begin{eqnarray}
\{Q_{A},Q_{B}\} &=& (\Delta_{\pm\mu})_{x\pm n_{\mu},x},
\label{Algebra_lat}
\end{eqnarray}
provided the following lattice Leibniz rule conditions hold:
\begin{eqnarray}
a_{A}+a_{B} &=& +n_{\mu} \ \ \ for \ \ \Delta_{+\mu}, \label{alg_forward}\\
a_{A}+a_{B} &=& -n_{\mu} \ \ \ for \ \ \Delta_{-\mu}. \label{alg_backward}
\end{eqnarray}
Figure \ref{s_Delta1} and \ref{s_Delta2} depict the possible
configurations of the general lattice SUSY algebra
(\ref{Algebra_lat}) subject to the conditions (\ref{alg_forward})
and (\ref{alg_backward}), respectively.
It is a non-trivial question to ask what type of SUSY algebras
satisfy these conditions. As described in \cite{DKKN1,DKKN2}, one
can show that the Dirac-K\"ahler twisted $N=D=2$ and $N=D=4$
satisfy the above conditions. Furthermore, it is recently shown in
\cite{DKKN3} that the twisted $N=4\ D=3$ algebra also satisfies
the conditions. We actually find the lattice realization of the
algebra (\ref{Alg_cont1})-(\ref{Alg_cont2}) as
\begin{eqnarray}
\{s,\overline{s}_{\mu}\} &\dot{=}& \Delta_{+\mu}, \hspace{39pt}
\{s_{\mu},\overline{s}_{\nu}\} \ \dot{=}\
\epsilon_{\mu\nu\rho}\Delta_{-\rho}, \label{Alg_lat1} \\ [2pt]
\{\overline{s},s_{\mu}\} &\dot{=}& -\Delta_{+\mu}, \hspace{20pt}
\{others\} \ =\ 0, \label{Alg_lat2}
\end{eqnarray}
where $\mu,\nu,\rho =1,2,3$ and the link anti-commutators in the
left hand side are understood. The corresponding Leibniz rule
conditions on the choice of $a_{A}$ can be expressed as
\begin{eqnarray}
a+\overline{a}_{\mu} &=& +n_{\mu}, \hspace{20pt}
a_{\mu}+\overline{a}_{\nu} \ =\ -|\epsilon_{\mu\nu\rho}|n_{\rho},
\hspace{20pt}
\overline{a}+a_{\mu} \ =\ +n_{\mu}, \label{Lattice_Leibniz}
\end{eqnarray}
which are satisfied by the following general solutions:
\begin{eqnarray}
a &=& (arbitrary), \hspace{20pt} \overline{a}_{\mu} \ =\ +n_{\mu} -a,
\label{const1} \\
a_{\mu} & =& -\sum_{\lambda\neq \mu}n_{\lambda} +a,
\hspace{20pt} \overline{a} \ =\ +\sum^{3}_{\lambda =1}n_{\lambda} -a.
\label{const2}
\end{eqnarray}
Note that there is a one-vector arbitrariness in the choice of
$a_{A}$, which eventually governs the resulting lattice
configuration of the model. We will come back to this point when
we construct the lattice Chern-Simons action. Notice also that the
total sum of all the shift parameters vanishes despite the
one-vector arbitrariness:
\begin{eqnarray}
\sum a_{A} &=& a+\overline{a}_{1}+\overline{a}_{2}+\overline{a}_{3}
+a_{1}+a_{2}+a_{3}+\overline{a} \ =\ 0.
\end{eqnarray}

\subsection{Twisted SUSY Chern-Simons multiplet on the lattice}

\indent

The lattice implementation of the twisted SUSY transformation laws
is possible only with an appropriate link assignment for each
component field. For example, the transformation law $sc=c^{2}$
requires that the ghost field $c$ should be located on a generic
link from $x$ to $x+a$ in order to be consistent with the link
assignment of $s$ which is also from $x$ to $x+a$. With this link
assignment, the corresponding lattice transformation law can be
expressed as
\begin{eqnarray}
(sc)_{x+2a,x} &=& (c)_{x+2a,x+a}(c)_{x+a,x}.
\end{eqnarray}
By studying all the twisted SUSY transformation laws in a similar
way, one finds that the link attributes can be consistently
assigned for all the component fields. Tables \ref{Shift} and
\ref{Trans_lat} summarize the link attributes of the component
fields and their twisted SUSY transformation laws. In Table
\ref{Shift} and in the following, the symbol $\sum n$ represents
the abbreviation $\sum_{\lambda =1}^{3}n_{\lambda}$. In Table
\ref{Trans_lat}, all the field products and (anti-)commutators
should be understood as link products and link (anti-)commutators,
with the link indices suppressed for simplicity. $D_{+\mu}$
denotes the covariant derivative with forward difference,
$D_{+\mu}\equiv \Delta_{+\mu}+A_{\mu}$.

\begin{table}
\begin{center}
\renewcommand{\arraystretch}{1.5}
\renewcommand{\tabcolsep}{5pt}
\begin{tabular}{c|cccc|cccc}
\hline
& $c$ & $\overline{c}$ & $A_{\mu}$ & $b$
& $s$ & $\overline{s}_{\mu}$ & $s_{\mu}$ & $\overline{s}$ \\ \hline
link & $(c)_{x+a,x}$ & $(\overline{c})_{x+\overline{a},x}$ & $(A_{\mu})_{x+n_{\mu},x}$
& $(b)_{x+\sum n, x}$
& $(s)_{x+a,x}$ & $(\overline{s}_{\mu})_{x+\overline{a}_{\mu},x}$
& $(s_{\mu})_{x+a_{\mu},x}$ & $(\overline{s})_{x+\overline{a},x}$.
\\ \hline
\end{tabular}
\caption{Link assignment of the fields and supercharges for a
generic value of $x$. Note that the shift parameters
$(a,\overline{a}_{\mu},a_{\mu},\overline{a})$ are subject to
(\ref{const1})-(\ref{const2}). } \label{Shift}
\end{center}
\end{table}

\begin{table}
\begin{center}
\renewcommand{\arraystretch}{1.4}
\renewcommand{\tabcolsep}{10pt}
\begin{tabular}{c|cccc}
\hline
& $s$ & $\overline{s}_{\rho}$ & $s_{\rho}$ & $\overline{s}$ \\ \hline
$c$ & $c^{2}$ & $-A_{\rho}$ & $0$ & $-b+\{\overline{c},c\}$\\
$\overline{c}$ & $b$ & $0$ & $A_{\rho}$ & $\overline{c}^{2}$ \\
$A_{\mu}$ & $-[D_{+\mu},c]$ & $-\epsilon_{\rho\mu\nu}[\Delta_{-\nu},\overline{c}]$ &
$-\epsilon_{\rho\mu\nu}[\Delta_{-\nu},c]$ & $-[D_{+\mu},\overline{c}]$ \\
$b$ & $0$ & $[\Delta_{+\rho},\overline{c}]$ & $[D_{+\rho},c]$
& $[\overline{c},b]$ \\ \hline
\end{tabular}
\caption{Twisted SUSY transformation laws on the lattice. The link
attributes of the products and (anti-)commutators are understood.}
\label{Trans_lat}
\end{center}
\end{table}

Notice that the gauge fields are associated only with the forward
difference and not with the backward difference. The absence of
the backward covariant derivative implies that the
(anti-)hermiticity can not be maintained if only one lattice
multiplet $(A_{\mu},b,c,\overline{c})$ is considered. One obvious
way to maintain the (anti-)hermiticity on the lattice is to
introduce the oppositely oriented multiplet associated with a set
of oppositely oriented supercharges. From now on, we slightly
change the notations and denote the set of supercharges introduced
in the above as
$s_{A}^{+}=(s^{+},\overline{s}_{\mu}^{+},s_{\mu}^{+},\overline{s}^{+})$.
Then we introduce an additional set of oppositely oriented
supercharges and denote them by
$s_{A}^{-}=(s^{-},\overline{s}_{\mu}^{-},s_{\mu}^{-},\overline{s}^{-})$.
The SUSY algebra is assumed to be
\begin{eqnarray}
\{s^{+},\os_{\mu}^{+}\} &\dot{=}& \Delta_{+\mu},   \hspace{20pt}
\{s_{\mu}^{+},\os_{\nu}^{+}\} \ \ \dot{=}\ \
\epsilon_{\mu\nu\rho}\Delta_{-\rho}, \hspace{20pt}
\{\os^{+},s_{\mu}^{+}\} \ \ \dot{=}\
-\Delta_{+\mu},\label{Algebra+} \\ [2pt] \{s^{-},\os_{\mu}^{-}\}
&\dot{=}& \Delta_{-\mu},  \hspace{20pt}
\{s_{\mu}^{-},\os_{\nu}^{-}\}  \ \ \dot{=}\ \
\epsilon_{\mu\nu\rho}\Delta_{+\rho}, \hspace{20pt}
\{\os^{-},s_{\mu}^{-}\} \ \ \dot{=}\ -\Delta_{-\mu},
\label{Algebra-}
\end{eqnarray}
with other anti-commutators of the supercharges vanishing:
$\{others\}=0$. We anticipate the on-shell closure of the algebra
and express them with dotted equalities. We have assumed that the
mixing sector of the algebra is just zero:
$\{s_{A}^{+},s_{B}^{-}\} = 0$. The hermitian conjugation of the
lattice supercharges and difference operators are defined as
\begin{eqnarray}
(s^{+})^{\dagger} &=& -s^{-},\hspace{30pt} (\os^{+})^{\dagger} \
=\ \os^{-}, \\ [2pt]
(s_{\mu}^{+})^{\dagger} &=& -s_{\mu}^{-},
\hspace{30pt}
(\os_{\mu}^{+})^{\dagger} \ =\ \os_{\mu}^{-},
\\ [2pt]
(\Delta_{+\mu})^{\dagger} &=& -\Delta_{-\mu}, \hspace{20pt}
(\Delta_{-\mu})^{\dagger} \ =\ -\Delta_{+\mu}.
\end{eqnarray}
We assign the supercharges $s_{A}^{+}$ and $s_{A}^{-}$ to be
located on the same links but with mutually opposite orientation,
namely $(s_{A}^{+})_{x+a_{A},x}$ and $(s_{A}^{-})_{x,x+a_{A}}$,
respectively, as summarized in Table \ref{Shift_comp}.
Correspondingly, we introduce oppositely oriented lattice
Chern-Simons multiplets $(c^{+},\oc^{+},A_{+\mu},b^{+})$ and
$(c^{-},\oc^{-},A_{-\mu},b^{-})$, together with the following
hermitian conjugation conditions,
\begin{eqnarray}
(c^{+})^{\dagger} &=& -c^{-}, \hspace{30pt} (\oc^{+})^{\dagger} \
=\ \oc^{-}, \label{hermiticily_comp1} \\ [2pt]
(A_{+\mu})^{\dagger} &=& -A_{-\mu}, \hspace{20pt}
(b^{+})^{\dagger} \ =\ -b^{-}. \label{hermiticily_comp2}
\end{eqnarray}
The link attributes of the multiplets and their SUSY
transformation laws are given in Tables \ref{Shift_comp} and
\ref{Trans}, respectively. The covariant differences in Table
\ref{Trans}, $D_{+\mu}$ and $D_{-\mu}$, are defined as $D_{\pm
\mu} \equiv \Delta_{\pm\mu}+A_{\pm\mu}$, which obey the obvious
hermitian conjugation relations, $D_{\pm\mu}^{\dagger} =
-D_{\mp\mu}$. We again assume that the SUSY transformation between
the different sectors be trivial, namely,
\begin{eqnarray}
[s_{A}^{+},\varphi^{-}\} &=& [s_{A}^{-},\varphi^{+}\} \ =\ 0,
\label{trivial}
\end{eqnarray}
where $\varphi^{+}$ and $\varphi^{-}$ denote any component of
 $(c^{+},\oc^{+},A_{+\mu},b^{+})$ and
$(c^{-},\oc^{-},A_{-\mu},b^{-})$, respectively. Note that although
the number of total supercharges is doubled in the present
(anti-)hermitian formulation, the lattice Leibniz rule
requirements associated with the algebra
(\ref{Algebra+})-(\ref{Algebra-}) remain unchanged and are
expressed as (\ref{Lattice_Leibniz}). The generic solutions are
still given by (\ref{const1})-(\ref{const2}).

\begin{table}
\renewcommand{\arraystretch}{1.5}
\renewcommand{\tabcolsep}{3pt}
\begin{tabular}{c|cccc|cccc}
\hline
& $s^{+}$ & $\os_{\mu}^{+}$ & $s_{\mu}^{+}$ & $\os^{+}$
& $s^{-}$ & $\os_{\mu}^{-}$ & $s_{\mu}^{-}$ & $\os^{-}$ \\ \hline
link
& $(s^{+})_{x+a,x}$ & $(\overline{s}_{\mu}^{+})_{x+\overline{a}_{\mu},x}$
& $(s_{\mu}^{+})_{x+a_{\mu},x}$ & $(\os^{+})_{x+\oa, x}$
& $(s^{-})_{x,x+a}$ & $(\overline{s}_{\mu}^{-})_{x,x+\overline{a}_{\mu}}$
& $(s_{\mu}^{-})_{x,x+a_{\mu}}$ & $(\os^{-})_{x, x+\oa}$  \\ \hline
\end{tabular}

\vspace{10pt}
\renewcommand{\arraystretch}{1.5}
\renewcommand{\tabcolsep}{3pt}
\begin{tabular}{c|cccc|cccc}
\hline
& $c^{+}$ & $\overline{c}^{+}$ & $A_{+\mu}$ & $b^{+}$
& $c^{-}$ & $\overline{c}^{-}$ & $A_{-\mu}$ & $b^{-}$
\\ \hline
link
& $(c^{+})_{x+a,x}$ & $(\overline{c}^{+})_{x+\overline{a},x}$
& $(A_{+\mu})_{x+n_{\mu},x}$
& $(b^{+})_{x+\sum n, x}$
& $(c^{-})_{x,x+a}$ & $(\overline{c}^{-})_{x,x+\overline{a}}$
& $(A_{-\mu})_{x,x+n_{\mu}}$
& $(b^{-})_{x, x+\sum n}$
\\ \hline
\end{tabular}
\caption{Link properties of oppositely oriented supercharges and
component fields}
\label{Shift_comp}
\end{table}

\begin{table}
\begin{center}
\renewcommand{\arraystretch}{1.4}
\renewcommand{\tabcolsep}{10pt}
\begin{tabular}{c|cccc}
\hline
& $s^{\pm}$ & $\overline{s}_{\rho}^{\pm}$ & $s_{\rho}^{\pm}$ & $\overline{s}^{\pm}$
\\ \hline
$c^{\pm}$ & $(c^{\pm})^{2}$ & $-A_{\pm\rho}$ & $0$
& $-b^{\pm}+\{\overline{c}^{\pm},c^{\pm}\}$
\\
$\overline{c}^{\pm}$ & $b^{\pm}$ & $0$ & $A_{\pm\rho}$ & $(\overline{c}^{\pm})^{2}$
\\
$A_{\pm\mu}$ & $-[D_{\pm\mu},c^{\pm}]$
& $-\epsilon_{\rho\mu\nu}[\Delta_{\mp\nu},\overline{c}^{\pm}]$ &
$-\epsilon_{\rho\mu\nu}[\Delta_{\mp\nu},c^{\pm}]$ & $-[D_{\pm\mu},\overline{c}^{\pm}]$
\\
$b^{\pm}$ & $0$ & $[\Delta_{\pm\rho},\overline{c}^{\pm}]$ & $[D_{\pm\rho},c^{\pm}]$
& $[\overline{c}^{\pm},b^{\pm}]$
\\ \hline
\end{tabular}
\caption{Twisted SUSY transformation laws on the lattice. The
upper and lower signs show the transformation laws of
$(c^{+},\overline{c}^{+},A_{+\mu},b^{+})$ under
$(s^{+},\overline{s}_{\mu}^{+},s_{+\mu},\overline{s})$ and
$(c^{+},\overline{c}^{+},A_{+\mu},b^{+})$ under
$(s^{+},\overline{s}_{\mu}^{+},s_{+\mu},\overline{s})$,
respectively. The link attributes of the products and
(anti-)commutators are understood.} \label{Trans}
\end{center}
\end{table}


\section{Lattice Chern-Simons Action}
\indent

In terms of the two oppositely oriented multiplets, the
anti-hermitian, Landau gauge fixed Chern-Simons action on a three
dimensional lattice is given by
\begin{eqnarray}
S^{tot} &=& k^{+}S^{+} + k^{-}S^{-}, \label{action_tot}
\end{eqnarray}
with
\begin{eqnarray}
S^{+} &=&  \frac{i}{4\pi} \sum_{x} \mathrm{Tr} \biggl[
\frac{1}{2}\epsilon_{\mu\nu\rho} (A_{+\mu})_{x+\sum n,
x+n_{\nu}+n_{\rho}}
[\Delta_{+\nu},A_{+\rho}]_{x+n_{\nu}+n_{\rho},x} \nonumber \\
[2pt] &&+\frac{1}{3}\epsilon_{\mu\nu\rho}
(A_{+\mu}A_{+\nu}A_{+\rho})_{x+\sum n ,x}
 - (b^{+})_{x+\sum n,x} [\Delta_{-\mu},A_{+\mu}]_{x,x} \nonumber \\ [2pt]
&&- (\overline{c}^{+})_{x+\sum n, x+a}
[\Delta_{-\mu},[D_{+\mu},c^{+}]]_{x+a,x} \biggr],
\label{Action+}
\\ [5pt]
S^{-} &=&  \frac{i}{4\pi} \sum_{x} \mathrm{Tr} \biggl[
\frac{1}{2}\epsilon_{\mu\nu\rho} (A_{-\mu})_{x-\sum n,
x-n_{\nu}-n_{\rho}}
[\Delta_{-\nu},A_{-\rho}]_{x-n_{\nu}-n_{\rho},x} \nonumber \\
[2pt] &&+\frac{1}{3}\epsilon_{\mu\nu\rho}
(A_{-\mu}A_{-\nu}A_{-\rho})_{x-\sum n ,x}
 - (b^{-})_{x-\sum n,x} [\Delta_{+\mu},A_{-\mu}]_{x,x} \nonumber \\ [2pt]
&&- (\overline{c}^{-})_{x-\sum n, x-a}
[\Delta_{+\mu},[D_{-\mu},c^{-}]]_{x-a,x} \biggr],
\label{Action-}
\end{eqnarray}
where $k^{+}$ and $k^{-}$ denote complex parameters related to
each other by complex conjugation, $(k^{\pm})^{*}=  k^{\mp}$. The
summation over $x$ in (\ref{Action+}) and (\ref{Action-}) covers
all the integer sites of a three dimensional regular lattice,
anticipating the fact that the $a$ needs to be integer vectors.
The anti-hermiticity of the total action is manifest.

\subsection{Twisted SUSY invariance}

\indent

\begin{figure}
\begin{center}
\begin{minipage}{70mm}
\begin{center}
\includegraphics[width=70mm]{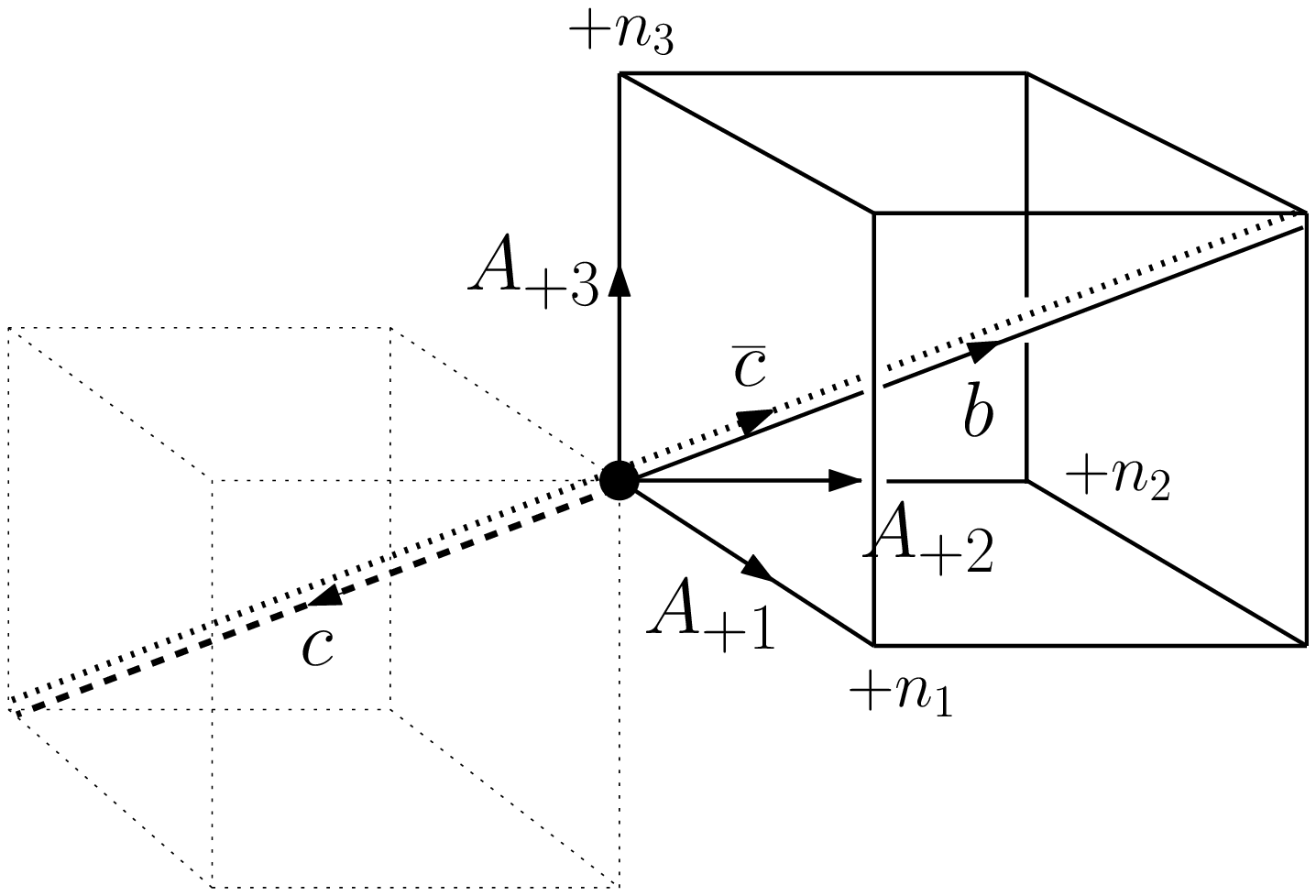}
\caption{Configurations of the component fields $(A_{+\mu},c,\overline{c},b)$
in the action $S^{+}$ for $a=-\sum n$ :
All the edges of each unit cell are occupied by $A_{+\mu}$.}
\label{CSfig1}
\end{center}
\end{minipage}
\hspace{10pt}
\begin{minipage}{70mm}
\begin{center}
\includegraphics[width=40mm]{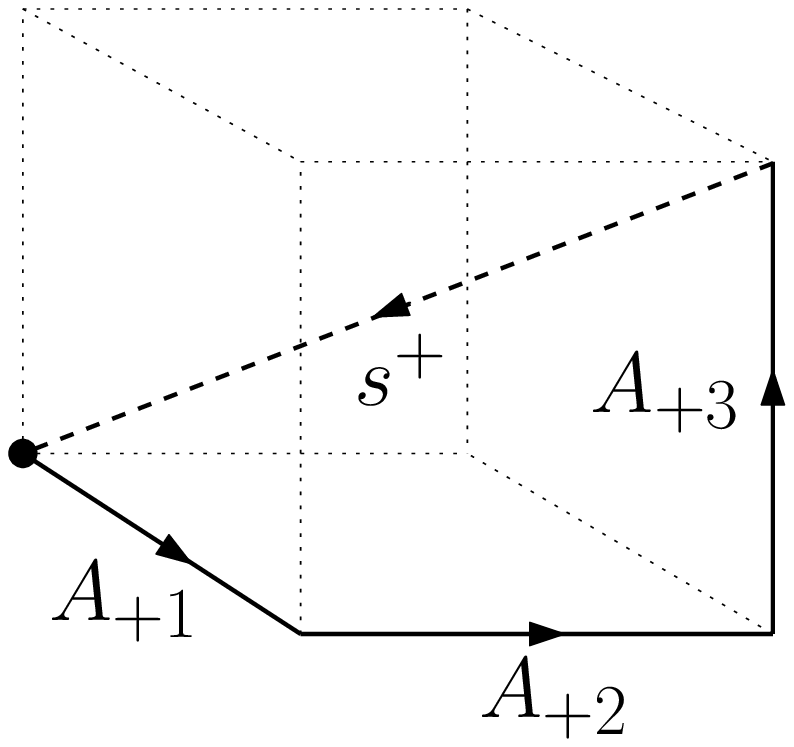}
\caption{A typical configuration in the transformed action $s^{+}S^{+}$}
\label{CSfig2}
\end{center}
\end{minipage}
\end{center}
\end{figure}

Before showing the SUSY invariance of the lattice action
(\ref{action_tot})-(\ref{Action-}), we would like to make the
following remarks. First, in order to ensure the SUSY invariance
of the action, one needs to take care of the ordering of the link
fields. The notion of proper ordering in lattice SUSY formulations
has been addressed in Ref. \cite{DKKN3}. Here in the lattice
Chern-Simons action, the proper ordering is nothing but the
geometrically connected ordering; namely, each term of $S^{+}$ or
$S^{-}$ consist of factors on connected links. Furthermore, all
the terms in $S^{+}$ and $S^{-}$ connect $x$ to $x+\sum n$ and $x$
to $x-\sum n$, respectively, through a sequence of links. The
homogeneous connecting property is a direct consequence of the
link component fields consistently allocated with the $N=4\ D=3$
twisted SUSY transformation laws on the lattice. Figure
\ref{CSfig1} depicts the configuration of the component fields per
unit cell in the case of $a=-\sum n$. The second remark is that
the $N=4\ D=3$ twisted SUSY invariance of the action
(\ref{action_tot}) is intrinsically related to the one-vector
arbitrariness (\ref{const1})-(\ref{const2}) in the solutions for
the lattice Leibniz rule conditions. Since the twisted SUSY
variations satisfy eq. (\ref{trivial}), the only non-trivial
variations come from either $s^{+}_{A}S^{+}$ or $s^{-}_{A}S^{-}$,
whose link attributes are given by $(x+\sum n +a_{A},x)$ and
$(x-\sum n -a_{A},x)$, respectively. One observes here that if one
takes $a_{A}= -\sum n$, then the twisted SUSY variation of the
action is reduced to that for closed loops.

The twisted SUSY invariance of the action can be explicitly
verified by exploiting the above remarks. For example, the $s^{+}$
variation of the second term in (\ref{Action+}) gives
\begin{eqnarray}
s^{+} {S}^{+}|_{2nd\ term} &=&  \frac{i}{4\pi}\sum_{x} \mathrm{Tr}
\biggl[
\frac{1}{3}\epsilon_{\mu\nu\rho}((s^{+}A_{+\mu})A_{+\nu}A_{+\rho})_{x+\sum
n +a,x} \nonumber \\ [2pt]
&&+\frac{1}{3}\epsilon_{\mu\nu\rho}(A_{+\mu}(s^{+}A_{+\nu})A_{+\rho})_{x+\sum
n +a,x} \nonumber \\ [2pt]
&&+\frac{1}{3}\epsilon_{\mu\nu\rho}(A_{+\mu}A_{+\nu}(s^{+}A_{+\rho}))_{x+\sum
n +a,x} \biggr],
\end{eqnarray}
whose link attribute is given by $(x+\sum n +a,x)$. If we take
$a=-\sum n$, then each term above is reduced to connected links
forming a closed loop. After using the cyclic property of trace
under the summation over $x$ and the $s^{+}$ transformation law of
$A_{+\mu}$, $s^{+}A_{+\mu}=-[D_{+\mu},c^{+}]$, one obtains
\begin{eqnarray}
s^{+} {S}^{+}|_{2nd\ term}
&=& - \frac{i}{4\pi} \sum_{x} \mathrm{Tr}\
\epsilon_{\mu\nu\rho}([D_{+\mu},c^{+}]A_{+\nu}A_{+\rho})_{x,x} \nonumber \\
&=& - \frac{i}{4\pi} \sum_{x} \mathrm{Tr} \biggl[
\epsilon_{\mu\nu\rho}([\Delta_{+\mu},c^{+}]A_{+\nu}A_{+\rho})_{x,x}
+\epsilon_{\mu\nu\rho}([A_{+\mu},c^{+}]A_{+\nu}A_{+\rho})_{x,x}
\biggr] \nonumber \\ [2pt] &=& - \frac{i}{4\pi} \sum_{x}
\mathrm{Tr}\
\epsilon_{\mu\nu\rho}([\Delta_{+\mu},c^{+}]A_{+\nu}A_{+\rho})_{x,x},
\end{eqnarray}
where from the first to the second line, we just inserted the
expression of forward covariant differences,
$D_{+\mu}=\Delta_{+\mu}+A_{+\mu}$, while from the second to the
third line, we used the trace property and anti-symmetric property
of $\epsilon_{\mu\nu\rho}$ to cancel out the second term. Figure
\ref{CSfig2} depicts the typical configuration of component fields
in the $s^{+}$ transformed action with the particular choice of
$a=-\sum n$. The operation of $s^{+}$ plays the role to close the
loop. Performing the same procedure for the other terms in
(\ref{Action+}), one can explicitly show that $s^{+}$ variations
of $S^{+}$ give the total difference terms which are vanishing
under the summation over $x$.
Furthermore, $s^{-}S^{-}=0$ can also be shown explicitly with the
choice of $a=-\sum n$. In a similar manner, we may verify the
invariance of the total action (\ref{action_tot}) with respect to
each supercharge of
$(s^{\pm},\overline{s}_{\mu}^{\pm},s_{\mu}^{\pm},\overline{s}^{\pm})$
under an appropriate choice of $a_{A}$:
\begin{eqnarray}
s^{\pm}S^{tot}&=&0 \ \ \ for \ \  a=-\sum n,  \label{SUSY_inv_1} \\
\overline{s}^{\pm}_{\mu}S^{tot}&=&0 \ \ \  for \ \
\overline{a}_{\mu}=-\sum n, \ \ (\mu =1,2,3) \label{SUSY_inv_2} \\
s^{\pm}_{\mu}S^{tot}&=&0 \ \ \  for \ \
a_{\mu}=-\sum n, \ \ (\mu=1,2,3) \label{SUSY_inv_3}\\
\overline{s}^{\pm}S^{tot}&=&0 \ \ \ for \ \ \overline{a}=-\sum n.
\label{SUSY_inv_4}
\end{eqnarray}
Notice again that the one-vector arbitrariness associated with the
lattice algebra (\ref{Algebra+})-(\ref{Algebra-}) has played a
fundamental role in the natural realization of the invariance
under the full lattice SUSY algebra.

Keeping in accordance with the above invariance of the lattice
Chern-Simons action, one may define the twisted SUSY variations
$\delta_{A}$ for the component fields as follows:
\begin{eqnarray}
\delta_{A}(\varphi)_{x+a_{\varphi},x} &=&
\overleftarrow{T}_{a_{A}}\eta_{A}(s_{A}
\varphi)_{x+a_{\varphi}+a_{A},x},\hspace{20pt}
\mathrm{(no\ sum)}
\label{variations}
\end{eqnarray}
where $(\varphi)_{x+a_{\varphi},x}$ denotes any of the component
fields $(c^{\pm},\overline{c}^{\pm},A_{\pm \mu},b^{\pm})$.
$\overleftarrow{T}_{a_{A}}$ represents a shift operator acting on
the functions from the right,
$f(x)\overleftarrow{T}_{a_{A}}=f(x+a_{A})$, while $\eta_{A}$
represents a constant Grassmann parameter. One can verify the
invariance of the total action (\ref{action_tot}) under the above
component-wise twisted SUSY variations,
\begin{eqnarray}
S^{tot}[\varphi+\delta_{A}\varphi] - S^{tot}[\varphi] &=& 0,
\hspace{20pt}
\mathrm{for} \ \ \
a_{A}= -\sum n,
\end{eqnarray}
where $\varphi$ represents collectively the set of all component
fields that appear in the total action. The existence of the shift
operator $\overleftarrow{T}_{a_{A}}$ in the component-wise SUSY
variations $(\ref{variations})$ and the notion of the proper
ordering in the lattice action seems to imply that the entire
lattice SUSY formulation could be embedded in a certain
non-commutative (super)space framework, which will be addressed in
the future development.

\subsection{Kernels}

\indent

Another important feature of the lattice Chern-Simons action
(\ref{action_tot})-(\ref{Action-}) is that the kinetic terms of
the lattice gauge fields can be expressed in terms of the
one-sided difference version of the Fr\"ohlich-Marchetti kernels
\cite{Frohlich},
\begin{eqnarray}
S^{+}|_{1st\ term} &=& \frac{i}{8\pi} \sum_{xy} \mathrm{Tr}\
(A_{+\mu})_{x+n_{\mu},x}\hat{K}_{\mu\nu}(x-y)(A_{+\nu})_{y+n_{\nu},y},
\label{1st+} \\ [2pt]
S^{-}|_{1st\ term} &=& \frac{i}{8\pi}
\sum_{xy} \mathrm{Tr}\ (A_{-\mu})_{x,x+n_{\mu}}
K_{\mu\nu}(x-y)(A_{-\nu})_{y,y+n_{\nu}}, \label{1st-}
\end{eqnarray}
where the kernels $K(x-y)$ and  $\hat{K}(x-y)$ are given by \cite{Berruto},
\begin{eqnarray}
K_{\mu\nu}(x-y) &=& \overrightarrow{T}_{n_{\mu}}
\epsilon_{\mu\rho\nu}\ \partial_{+\rho}\ \delta_{xy}, \label{kernel1}
\\ [2pt]
\hat{K}_{\mu\nu}(x-y) &=& \overrightarrow{T}_{-n_{\nu}}
\epsilon_{\mu\rho\nu}\ \partial_{-\rho}\ \delta_{xy}, \label{kernel2}
\end{eqnarray}
with $\partial_{+\mu}f(x) = f(x+n_{\mu})-f(x)$,
$\partial_{-\mu}f(x) = f(x)-f(x-n_{\mu})$,
$\overrightarrow{T}_{n_{\mu}}f(x) = f(x+n_{\mu})$ and
$\overrightarrow{T}_{-n_{\mu}}f(x) \ =\ f(x-n_{\mu})$. Since the
gauge fields are located on links, the analysis in the momentum
space superficially depends on where to pick up their
representatives in the configuration space. Fourier transformation
of the link gauge fields is given by
\begin{eqnarray}
(A_{+\mu})_{x+n_{\mu},x} &=& \int_{B} \frac{d^{3}p}{(2\pi)^{3}}
e^{-ip\cdot (x+\alpha n_{\mu})} A_{+\mu}(p), \\
(A_{-\mu})_{x,x+n_{\mu}} &=& \int_{B} \frac{d^{3}p}{(2\pi)^{3}}
e^{-ip\cdot (x+\alpha n_{\mu})} A_{-\mu}(p),
\end{eqnarray}
where the constant $\alpha$ parameterizes the representative
points of the gauge fields. Namely $\alpha = 0$, $\frac{1}{2}$ and
$1$ correspond to the initial point, mid point and ending point
prescriptions, respectively. $B$ denotes the Brillouin zone:
$B=\{p_{\mu}|-\pi \leq p_{\mu} \leq \pi, \mu =1,2,3\}$.
$A_{+\mu}(p)$ and $A_{-\mu}(p)$ are related to each other by the
complex conjugation, $A_{\pm\mu}(p)^{\dagger}=-A_{\mp\mu}(-p)$, in
order to satisfy the conjugation relation of the gauge fields in
the configuration space. Momentum space representation of the
kernels (\ref{kernel1}) and (\ref{kernel2}) is accordingly given
by
\begin{eqnarray}
K^{(\alpha)}_{\mu\nu}(p) &=& -2i \epsilon_{\mu\rho\nu}\
e^{-i((1-\alpha) p_{\mu} + \alpha p_{\nu} + \frac{1}{2}p_{\rho})}
\sin\frac{p_{\rho}}{2}, \label{kernel3} \\ [2pt]
\hat{K}^{(\alpha)}_{\mu\nu}(p) &=& -2i \epsilon_{\mu\rho\nu}\
e^{+i(\alpha p_{\mu} + (1-\alpha)p_{\nu} + \frac{1}{2}p_{\rho})}
\sin\frac{p_{\rho}}{2}. \label{kernel4}
\end{eqnarray}
Although the form of the kernels is explicitly dependent on the
parameter $\alpha$, their eigenvalues should be independent of
$\alpha$. In fact, one may easily verify that the eigenvalues of
$K$ are given by $\lambda(p)= 0, \pm
2e^{-\frac{i}{2}\sum_{\mu=1}^{3}p_{\mu}}
\sqrt{\sum_{\mu=1}^{3}\sin^{2}\frac{p_{\mu}}{2}}$. Likewise the
eigenvalues of $\hat{K}$ are given by $\hat{\lambda}(p)= 0, \pm
2e^{+\frac{i}{2}\sum_{\mu=1}^{3}p_{\mu}}
\sqrt{\sum_{\mu=1}^{3}\sin^{2}\frac{p_{\mu}}{2}}$, the complex
conjugate of $\lambda(p)$. The zero eigenvalue, which arises from
the original gauge invariance of the action, should be cured by
the gauge-fixing terms. It is important to notice that they do not
have any other extra zero eigenvalues, which implies that both of
(\ref{kernel3}) and (\ref{kernel4}) could serve as the invertible
kernels after the gauge-fixing terms are properly taken into
account. Notice again that the eigenvalues always come in complex
conjugated pairs, ensuring the anti-hermiticity of the entire
formulation. These features are direct consequences of the use of
two sets of oppositely oriented component fields on the lattice.

\subsection{Naive continuum limit}

\indent

The naive continuum limit of the total action is taken by
replacing the difference operators by differential operators,
\begin{eqnarray}
\Delta_{\pm\mu} &\rightarrow& \partial_{\mu}.
\end{eqnarray}
The hermitian conjugation property of $\Delta_{\pm\mu}$ is
accordingly reduced into the anti-hermiticity of $\partial_{\mu}$,
\begin{eqnarray}
(\Delta_{\pm\mu})^{\dagger} \ =\ -\Delta_{\mp\mu}
& \rightarrow &
(\partial_{\mu})^{\dagger} \ =\ -\partial_{\mu}.
\end{eqnarray}
The hermitian conjugation properties, eqs.
(\ref{hermiticily_comp1})-(\ref{hermiticily_comp2}), of the
component fields are supposed to be retained in the continuum. The
component fields in the continuum limit are accordingly given by
\begin{eqnarray}
(A_{+\mu})_{x+n_{\mu},x} &\rightarrow& A^{+}_{\mu}(x)\equiv
A_{\mu}(x)+iB_{\mu}(x), \hspace{5pt} (A_{-\mu})_{x,x+n_{\mu}}
\rightarrow A^{-}_{\mu}(x)\equiv A_{\mu}(x)-iB_{\mu}(x), \qquad
\label{cont_A} \\ [2pt] (c^{+})_{x+a,x} &\rightarrow& c^{+}(x)
\equiv c(x) + i d(x), \hspace{37pt} (c^{-})_{x,x+a} \ \rightarrow\
c^{-}(x) \equiv c(x) - i d(x), \\ [2pt]
(\overline{c}^{+})_{x+\oa,x} &\rightarrow&
\overline{c}^{+}(x)\equiv \overline{c}(x) + i \overline{d}(x),
\hspace{37pt} (\overline{c}^{-})_{x,x+\oa} \ \rightarrow\
\overline{c}^{-}(x)\equiv \overline{c}(x) - i \overline{d}(x),
\\ [2pt]
(b^{+})_{x+\sum n, x} &\rightarrow& b^{+}(x) \equiv b(x) + i h(x),
\hspace{26pt} (b^{-})_{x, x+\sum n} \ \rightarrow\ b^{-}(x) \equiv
b(x) - i h(x).
\end{eqnarray}
Here $(A_{\mu},B_{\mu},b,h)$, $(c,d)$ and
$(\overline{c},\overline{d})$ denote bosonic anti-hermitian
fields, Grassmann odd anti-hermitian fields and Grassmann odd
hermitian fields, respectively. Note that the two possible
orientations of the lattice component fields can naturally be
interpreted as the complex structure of the gauge group. In terms
of the above expansions, the entire action
(\ref{action_tot})-(\ref{Action-}) can be expressed in the
continuum limit as
\begin{eqnarray}
S^{tot}_{cont} &=& k^{+} S^{+}_{cont} + k^{-}S^{-}_{cont}
\nonumber \\ [4pt]
&=& \frac{i}{2\pi} u \int d^{3}x \ \mathrm{Tr}\
\biggl[ \frac{1}{2}\epsilon_{\mu\nu\rho}
(A_{\mu}\partial_{\nu}A_{\rho} - B_{\mu}\partial_{\nu}B_{\rho})
+\frac{1}{3}\epsilon_{\mu\nu\rho}
(A_{\mu}A_{\nu}A_{\rho}-3A_{\mu}B_{\nu}B_{\rho}) \nonumber \\
[0pt] &&-b \partial_{\mu}A_{\mu} + h\partial_{\mu}B_{\mu} -
\overline{c}\partial_{\mu}(D_{\mu}c -[B_{\mu},d]) +
\overline{d}\partial_{\mu}(D_{\mu}d +[B_{\mu},c]) \biggr]
\nonumber
\\ [4pt]
&&- \frac{i}{2\pi} v \int d^{3}x \ \mathrm{Tr}\ \biggl[
\frac{1}{2}\epsilon_{\mu\nu\rho} (A_{\mu}\partial_{\nu}B_{\rho} +
B_{\mu}\partial_{\nu}A_{\rho}) +\frac{1}{3}\epsilon_{\mu\nu\rho}
(3A_{\mu}A_{\nu}B_{\rho}-B_{\mu}B_{\nu}B_{\rho}) \nonumber
\\ [0pt]
&&-b \partial_{\mu}B_{\mu} -h \partial_{\mu}A_{\mu} -
\overline{d}\partial_{\mu}(D_{\mu}c -[B_{\mu},d]) -
\overline{c}\partial_{\mu}(D_{\mu}d +[B_{\mu},c]) \biggr] ,
\label{action_cont}
\end{eqnarray}
where the constants $u$ and $v$ are the real and imaginary part of
the complex parameters $k^{\pm} = u \pm iv$. The covariant
derivative $D_{\mu}$ is again defined by $D_{\mu}c =
\partial_{\mu}c+[A_{\mu},c]$. The action (\ref{action_cont}) can
be regarded as the Landau gauge fixed version of the Chern-Simons
action with complex gauge group originally proposed in Ref.
\cite{Witten:1989ip}. Obviously, if one takes
$B_{\mu}=d=\overline{d}=h=0$, the entire action
(\ref{action_cont}) is reduced into the expression (\ref{CSgf})
with the coefficient $u=k$. In the general case, according to ref.
\cite{Witten:1989ip}, the parameter $u$ must always be quantized
to be an integer $k$ if the $\mathrm{Tr}$ is normalized correctly,
while there is no quantization condition for the real parameter
$v$.

The lattice supercharges
$(s^{\pm},\os^{\pm}_{\mu},s^{\pm}_{\mu},\os^{\pm})$ may also be
expanded as
\begin{eqnarray}
s^{\pm}&=&\frac{1}{2}(s^{(1)}\pm is^{(2)}), \hspace{20pt}
\os^{\pm} \ =\ \frac{1}{2}(\os^{(1)}\mp i\os^{(2)}), \\
\os_{\mu}^{\pm} & =&  \frac{1}{2}(\os^{(1)}_{\mu}
\mp i\os^{(2)}_{\mu}), \hspace{20pt}
s_{\mu}^{\pm} \ =\ \frac{1}{2}(s_{\mu}^{(1)}\pm is_{\mu}^{(2)}),
\end{eqnarray}
with which the naive continuum limit of the lattice SUSY algebra
(\ref{Algebra+})-(\ref{Algebra-}) is given by
\begin{eqnarray}
\{s^{(i)},\os^{(j)}_{\mu}\} &\dot{=}& (\delta^{ij}\pm i \epsilon^{ij})\partial_{\mu},
\hspace{24pt}
\{s^{(i)}_{\mu},\os^{(j)}_{\nu}\} \ \dot{=}\ (\delta^{ij}\pm i \epsilon^{ij})
\epsilon_{\mu\nu\rho}\partial_{\rho}, \\
\{\os^{(i)},s^{(j)}_{\mu}\} &\dot{=}& -(\delta^{ij}\mp i \epsilon^{ij})\partial_{\mu},
\hspace{15pt}
\{others\} \ =\ 0,
\end{eqnarray}
for the continuum-limit multiplet
$\varphi^{\pm}=(c^{\pm},\oc^{\pm},A^{\pm}_{\mu},b^{\pm})$,
respectively. The suffixes $i,j$ take 1 or 2, and
$\epsilon^{12}=-\epsilon^{21}=1$. The SUSY transformation laws in
terms of $(s^{(i)},\os^{(i)}_{\mu},s^{(i)}_{\mu},\os^{(i)})$ for
the expanded component fields $(A_{\mu},B_{\mu},c,d,\oc,\od,b,h)$
are summarized in Table \ref{Trans_cont}. It is straightforward to
verify that the action with the coefficient $u$ and the action
with $v$ in (\ref{action_cont}) are separately invariant under the
twisted SUSY transformations
$(s^{(i)},\os^{(i)}_{\mu},s^{(i)}_{\mu},\os^{(i)})$.

\begin{table}[t]
\begin{center}
\renewcommand{\arraystretch}{1.2}
\renewcommand{\tabcolsep}{10pt}
\begin{tabular}{c|cccc}
\hline
& $s^{(1)}$ & $\os_{\rho}^{(1)}$ & $s_{\rho}^{(1)}$ & $\overline{s}^{(1)}$
\\ \hline
$c$ & $c^{2}-d^{2}$ & $-A_{\rho}$ & $0$
& $-b+\{\oc,c\}-\{\od,d\}$ \\
$d$ & $\{c,d\}$ & $-B_{\rho}$ & $0$

& $-h+\{\oc,d\}+\{\od,c\}$ \\
$\oc$ & $b$ & $0$ & $A_{\rho}$
& $\oc^{2}-\od^{2}$
\\
$\od$ & $h$ & $0$ & $B_{\rho}$
& $\{\oc,\od\}$
\\
$A_{\mu}$ & $-D_{\mu}c+[B_{\mu},d]$
& $-\epsilon_{\rho\mu\nu}\partial_{\nu}\oc$ &
$-\epsilon_{\rho\mu\nu}\partial_{\nu}c$
& $-D_{\mu}\oc+[B_{\mu},\od]$
\\
$B_{\mu}$ & $-D_{\mu}d-[B_{\mu},c]$
& $-\epsilon_{\rho\mu\nu}\partial_{\nu}\od$ &
$-\epsilon_{\rho\mu\nu}\partial_{\nu}d$
& $-D_{\mu}\od-[B_{\mu},\oc]$
\\
$b$ & $0$ & $\partial_{\rho}\oc$
& $D_{\rho}c-[B_{\rho},d]$
& $[\oc,b]-[\od,h]$
\\
$h$ & $0$ & $\partial_{\rho}\od$
& $D_{\rho}d+[B_{\rho},c]$
& $[\oc,h]+[\od,b]$
\\
\hline
\end{tabular}
\vspace{5pt}
\renewcommand{\arraystretch}{1.2}
\renewcommand{\tabcolsep}{10pt}
\begin{tabular}{c|cccc}
\hline
& $s^{(2)}$ & $\os_{\rho}^{(2)}$ & $s_{\rho}^{(2)}$ & $\overline{s}^{(2)}$
\\ \hline
$c$ & $\{c,d\}$ & $B_{\rho}$ & $0$
& $h-\{\oc,d\}-\{\od,c\}$ \\
$d$ & $-c^{2}+d^{2}$ & $-A_{\rho}$ & $0$
& $-b+\{\oc,c\}-\{\od,d\}$ \\
$\oc$ & $h$ & $0$ & $B_{\rho}$
& $-\{\oc,\od\}$
\\
$\od$ & $-b$ & $0$ & $-A_{\rho}$
& $\oc^{2}-\od^{2}$
\\
$A_{\mu}$ & $-D_{\mu}d-[B_{\mu},c]$
& $\epsilon_{\rho\mu\nu}\partial_{\nu}\od$ &
$-\epsilon_{\rho\mu\nu}\partial_{\nu}d$
& $D_{\mu}\od+[B_{\mu},\oc]$
\\
$B_{\mu}$ & $D_{\mu}c-[B_{\mu},d]$
& $-\epsilon_{\rho\mu\nu}\partial_{\nu}\oc$ &
$\epsilon_{\rho\mu\nu}\partial_{\nu}c$
& $-D_{\mu}\oc+[B_{\mu},\od]$
\\

$b$ & $0$ & $-\partial_{\rho}\od$
& $D_{\rho}d+[B_{\rho},c]$
& $-[\oc,h]-[\od,b]$
\\
$h$ & $0$ & $\partial_{\rho}\oc$
& $-D_{\rho}c+[B_{\rho},d]$
& $[\oc,b]-[\od,h]$
\\
\hline
\end{tabular}
\caption{Twisted SUSY transformation laws
in the naive continuum limit
for the expanded component fields $(c,d,\oc,\od,A_{\mu},B_{\mu},b,h)$}
\label{Trans_cont}
\end{center}
\end{table}

\section{Transformation Properties under Parity}

\indent

The properties under parity transformation are an important issue
for continuum Chern-Simons theory. In this section we address this
issue for our twisted SUSY Chern-Simons action on a lattice. We
first recall that on a Euclidean three dimensional lattice or
spacetime, parity may be defined by the simultaneous inversion of
all coordinates (\ref{def_parity}). Since the gauge fields $A_{\pm
\mu}$ are located on links $(A_{\pm \mu})_{x\pm n_{\mu},x}$ and
the parity also flips the link orientations, one may naturally
define the parity operation $P$ for $A_{\pm \mu}$ on the lattice
by
\begin{eqnarray}
P (A_{+\mu})_{x+n_{\mu},x} P^{-1}
&=& - (A_{-\mu})_{-x-n_{\mu},-x}
\label{parityAmu}
\end{eqnarray}
where $-x$ denotes $(-x_{1},-x_{2},-x_{3})$. The difference
operators are also located on links so that their parity
transformation law is
\begin{eqnarray}
P (\Delta_{+\mu})_{x+n_{\mu},x} P^{-1} &=& -
(\Delta_{-\mu})_{-x-n_{\mu},-x} ,  \label{parityD}
\end{eqnarray}
which is also consistent with the fact that $\Delta_{\pm \mu}$
actually take the unit values in the link commutators (see
(\ref{delta=1})). As for the gauge-fixing component fields and the
supercharges, we define
\begin{eqnarray}
P (c^{+})_{x+a ,x} P^{-1} &=& + (c^{-})_{-x-a,-x}, \label{parityc}
\hspace{30pt} P (\overline{c}^{+})_{x+\overline{a} ,x} P^{-1} \ =\
- (\overline{c}^{-})_{-x-\overline{a},-x}, \label{parityoc}\\
[2pt] P (b^{+})_{x+\sum n,x} P^{-1} &=& - (b^{-})_{-x-\sum n,-x},
\label{parityb} \\ [4pt] P (s^{+})_{x+a,x} P^{-1} &=&
+(s^{-})_{-x-a,-x}, \hspace{29pt} P (\overline{s}^{+})_{x+\oa,x}
P^{-1} \ = \ -(\overline{s}^{-})_{-x-\oa,-x},
\label{paritys1}\\
P (\overline{s}_{\mu})_{x+\oa_{\mu},x} P^{-1}
&=& -(\overline{s}_{\mu})_{-x-\oa_{\mu},-x}, \hspace{20pt}
P (s_{\mu}^{+})_{x+a_{\mu},x} P^{-1} \ =\ +(s^{-}_{\mu})_{-x-a_{\mu},-x}.
\label{paritys2}
\end{eqnarray}
In the following we will see two interesting features resulting
from these definitions. One is regarding the parity of the lattice
Chern-Simons action. The other one is the parity property in the
continuum limit.

As for the parity transformation of the action, it is easy to see
that the definitions (\ref{parityAmu})-(\ref{parityb}) interchange
the two oppositely oriented parts of the action, $S^{+}$ given by
eq. (\ref{Action+}) and $S^{-}$ given by eq. (\ref{Action-}):
\begin{eqnarray}
P S^{+} P^{-1} &=&  \frac{i}{4\pi} P \sum_{x} \mathrm{Tr}\biggl[
\frac{1}{2}\epsilon_{\mu\nu\rho} (A_{+\mu})_{x+\sum n,
x+n_{\nu}+n_{\rho}} [\Delta_{+\nu},
A_{+\rho}]_{x+n_{\nu}+n_{\rho},x} + \cdots \biggr]
P^{-1} \nonumber \\
&=& -\frac{i}{4\pi}\sum_{x} \mathrm{Tr} \biggl[ \frac{1}{2}
\epsilon_{\mu\nu\rho} (A_{-\mu})_{-x-\sum n, -x-n_{\nu}-n_{\rho}}
[\Delta_{-\nu}, A_{-\rho}]_{-x-n_{\rho}-n_{\rho},-x} + \cdots
\biggr]
\nonumber  \\
&=& -\frac{i}{4\pi}\sum_{x} \mathrm{Tr} \biggl[ \frac{1}{2}
\epsilon_{\mu\nu\rho} (A_{-\mu})_{x-\sum n, x-n_{\nu}-n_{\rho}}
[\Delta_{-\nu}, A_{-\rho}]_{x-n_{\rho}-n_{\rho},x} + \cdots
\biggr] \nonumber  \\ [2pt] &=& -S^{-}.
\end{eqnarray}
Here from the second line to the third, we have replaced
$x\rightarrow -x$.
Likewise, we also have $P S^{-} P^{-1}=
-S^{+}$. We thus have the parity transformation for the total
action $S^{tot}$ (\ref{action_tot}) as
\begin{eqnarray}
P S^{tot} P^{-1} &=& - k^{+} S^{-} - k^{-}S^{+} ,
\end{eqnarray}
which implies that the total action is not an eigenstate of the
parity defined by (\ref{parityAmu})-(\ref{parityb}). Writing the
complex parameters $k^{\pm}$ as $k^{\pm}=u\pm iv$, we actually
have
\begin{eqnarray}
S^{tot} &=& u(S^{+}+S^{-}) + iv (S^{+}-S^{-}),
\\ [2pt]
P S^{tot}P^{-1} &=& - u(S^{+}+S^{-}) + iv (S^{+}-S^{-}).
\end{eqnarray}
Now it becomes clear that the total action is a sum of a parity
even part with the coefficient $u$ and a parity odd part with the
coefficient $iv$.
\begin{eqnarray}
 S^{tot}|_{v=0}  : \mathrm{parity \ odd},  \hspace{40pt}
 S^{tot}|_{u=0}  : \mathrm{parity \ even}.
\end{eqnarray}

One can understand the mixed behavior of the total action under
parity more clearly by examining the parity behavior of the
component fields in the continuum limit. In fact, by considering
the continuum limit (\ref{cont_A}) of the lattice parity operation
(\ref{parityAmu}), one obtains
\begin{eqnarray}
P A_{\mu}(x) P^{-1} &=& -A_{\mu}(-x), \label{parityA} \hspace{30pt}
P B_{\mu}(x) P^{-1} \ =\ +B_{\mu}(-x). \label{parityB}
\end{eqnarray}
which imply that the  $A_{\mu}(x)$ is an ordinary vector while the
$B_{\mu}(x)$ a pseudo-vector. By considering the continuum limit
of the relations (\ref{parityc})-(\ref{paritys2}), one also
obtains the parity behavior of the other component fields and the
supercharges as listed in Table {\ref{parity}}. In the language of
forms, the complex gauge fields $A^{\pm}_{\mu}$ may be regarded as
complex combinations of a one-form $A$ and a two-form $B$,
\begin{eqnarray}
A^{\pm}_{\mu}dx_{\mu} \ =\ A \pm i*B \label{form}
\end{eqnarray}
where $A=A_{\mu}dx_{\mu}$ and
$B=\frac{1}{2}B_{\mu\nu}dx_{\mu}\wedge dx_{\nu}$. The symbol $*$
denotes the Hodge star operation. Likewise, the continuum limit of
the gauge fixing component fields
$(c^{\pm},\overline{c}^{\pm},b^{\pm})$ are divided into the
complex combinations of 0-forms and 3-forms. It is interesting to
note that our anti-hermitian lattice formulation together with the
twisted SUSY structure actually involves all possible simplicial
forms in the three dimensional spacetime.

\begin{table}
\begin{center}
\renewcommand{\arraystretch}{1.4}
\renewcommand{\tabcolsep}{5pt}
\begin{tabular}{c|cccccccc|cccccccc|c}
\hline
& $A_{\mu}$ & $B_{\mu}$ & $c$ & $d$ & $\oc$ & $\od$
& $b$ & $h$
& $s^{(1)}$ & $s^{(2)}$ & $\os^{(1)}_{\mu}$ & $\os^{(2)}_{\mu}$
& $s^{(1)}_{\mu}$ & $s^{(2)}_{\mu}$ & $\os^{(1)}$ & $\os^{(2)}$
& $\partial_{\mu}$\\ \hline
parity & $-$ & $+$ & $+$ & $-$ & $-$ & $+$ & $-$ & $+$
& $+$ & $-$ & $-$ & $+$ & $+$ & $-$ & $-$ & $+$
& $-$ \\
\hline
\end{tabular}
\caption{Behavior under parity of the component fields and
supercharges in the continuum limit} \label{parity}
\end{center}
\end{table}

The mixed behavior under parity of the continuum action
(\ref{action_cont}) is now clearly understood. One can easily see
from Table \ref{parity} that part of the action with the
coefficient $u$ is actually parity odd, just like the ordinary
Chern-Simons action for a single gauge field, while part of the
action with the coefficient $v$ is parity even. The manifestly
anti-hermitian formulation on the lattice thus eventually leads to
a unified picture of even and odd parity Chern-Simons theory. It
is worthwhile to mention that the parity even part of the
continuum action (\ref{action_cont}) shares the same parity
behavior as the so-called ``dumbbell" Chern-Simons action
addressed in \cite{Kantor}, where vector and pseudo-vector gauge
fields are introduced as the lattice objects dual to each other.
We also note that the parity even part of the continuum action
(\ref{action_cont}) shares the same parity behavior with the
so-called ``doubled" Chern-Simons theory discussed in
\cite{Freedman03}, though the action is actually not the same.


\section{Summary \& Discussions}
\indent

We have constructed the Landau gauge fixed Chern-Simons theory on 
a three dimensional regular lattice. The $N=4\ D=3$ twisted SUSY
associated with the Chern-Simons action in Landau gauge has played
a crucial role as the guiding principle in the present lattice
construction. The one-vector arbitrariness associated with the 
$N=4\ D=3$ lattice algebra is shown to play an important role 
in maintaining the twisted SUSY invariance of the lattice action.
In order to ensure the manifest anti-hermiticity on the lattice,
we have introduced two sets of oppositely oriented component 
fields attached to every possible link. Owing to this ``doubling"
of the lattice component fields, the gauge kernels are shown to 
be free from the extra zero-eigenvalue problem. We have also
addressed the transformation properties under parity of the 
fields involved in our construction. It was pointed out that 
a natural definition of parity on the lattice involves component 
fields of opposite parity. Parity invariance then puts a constraint 
between the coefficients in front of the actions for the oppositely 
oriented component fields.

 
It is important to ask whether one can recover the appropriate 
$N=4\ D=3$ twisted SUSY Chern-Simons theory in the continuum 
limit. In particular, whether the continuum rotational symmetry 
and the entire $N=4\ D=3$ twisted SUSY invariance are restored 
in the continuum limit is an important issue, worth further 
study. Discussing these aspects requires a careful examination 
of possible quantum corrections on the lattice. Here we would 
like to point out an important correlation between the 
rotational symmetry and the twisted SUSY invariance of the 
lattice action (\ref{SUSY_inv_1})-(\ref{SUSY_inv_4}). 
Since in our formulation we respect only part of the entire
set of SUSY generators, not only the continuous rotational 
symmetry but also the discrete rotational symmetry (for the 
square lattice) are broken on the lattice. However, as one 
can see in Table \ref{Shift_comp} and Fig. \ref{CSfig1},
the lattice action with the parameter choice $a=-\sum n$, 
which corresponds to the invariance (\ref{SUSY_inv_1}),
has a symmetry subgroup with a single 3-fold rotation axis, 
$\mathcal{C}_{3}=(E,C_{3},C_{3}^{2})$, of the octahedral 
group $\mathcal{O}$. This is because all the gauge-fixing 
component fields $(c^{\pm},\overline{c}^{\pm},b^{\pm})$
are located on the diagonal link parallel to $a=-\sum n$,
while the gauge fields $A_{\pm\mu}$ are located on the 
regular edges.
The lattice action with $\overline{a}=-\sum n$,
which corresponds to the invariance (\ref{SUSY_inv_4}),
also has the same symmetry.
\begin{figure} 
\begin{center}
\includegraphics[width=40mm]{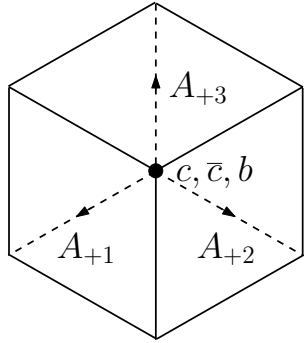}
\caption{Field configurations of Fig. \ref{CSfig1}
on a projected plane normal to $a=-\sum n$:
All the edges are occupied by $A_{+\mu}$.}
\label{Projected_plane}
\end{center}
\end{figure}
Fig. \ref{Projected_plane} shows the projected field 
configurations normal to $a=-\sum n$, where the 3-fold 
rotational symmetry is manifest. It is interesting to 
notice that the gauge-fixing component fields 
$(c,\overline{c},b)$ and the supercharge $s$
are projected onto a point, which corresponds to the 
fact that these component fields and the supercharge 
should behave as (pseudo-)scalars in the continuum limit.  

It should be stressed again that our lattice action is a 
gauge fixed one, nor really invariant under gauge symmetry.  
Instead, it has $N=4\ D=3$ twisted supersymmetry whose 
scalar transformation corresponds to the BRST transformation 
associated with the Landau-gauge fixed Chern-Simons theory. 
Although further study is needed to clarify whether the entire 
$N=4\ D=3$ twisted SUSY invariance can be properly restored 
in the continuum limit, the following three important features 
of our formulation may be explored to argue for the gauge 
invariance in the continuum limit: 1) The Landau gauge-fixed 
action (\ref{CSgf}) enables us to make use of the $N=4\ D=3$ 
twisted SUSY structure in building the lattice action; 
2) the remnant of the gauge symmetry in the original 
Chern-Simons action has turned into the scalar part of 
the $N=4\ D=3$ twisted SUSY; 3) the {\it infinitesimal} BRST 
transformations are preserved on the lattice. Therefore, 
at least formally, in our gauge fixed formulation there is 
no need to be concerned about large gauge transformations, 
which would be far more difficult to realize directly on 
the lattice. 


It is also important to ask whether the lattice formulation 
presented in this paper could really serve as a useful 
regularization scheme; namely, whether the quantum aspects 
such as the Chern-Simons coefficient renormalization 
\cite{CSW90} could be calculated in this framework. We 
should also address the possibility that the entire lattice 
SUSY description presented in this paper could be formulated 
more rigidly in terms of a certain non-commutative (super)space 
formalism. The work is in progress.

Another interesting question for possible applications in physics
is whether there exists a real or model system in condensed matter
physics that has a topological phase described by the Chern-Simons
action with complex gauge group.

\section*{Acknowledgments}

KN would like to thank N. Kawamoto for discussions and comments.
KN is supported by Department of Energy US Government, Grant No.
FG02-91ER 40661. YSW was supported in part by US NSF through grant
No. PHY-0457018.

\end{document}